\documentclass[a4paper]{article}

\usepackage{INTERSPEECH2022}
\usepackage{amssymb,amsmath,amsfonts}
\usepackage{physics}

\usepackage[caption=false,font=footnotesize]{subfig}
\usepackage{comment}
\usepackage{microtype}

\usepackage{booktabs}
\usepackage{cite}

\title{
Direction-Aware Joint Adaptation of Neural Speech Enhancement and Recognition 
in Real Multiparty Conversational Environments
}

\name{
Yicheng Du$^{1*}$\thanks{*These three authors contributed equally to this work.}, 
Aditya Arie Nugraha$^{2*}$,
Kouhei Sekiguchi$^{2*}$, \\
Yoshiaki Bando$^{3,2}$, 
Mathieu Fontaine$^{4,2}$, 
Kazuyoshi Yoshii$^{1,2}$
}
\address{
  $^1$Graduate School of Informatics, Kyoto University, Japan \ \ \ $^2$AIP, RIKEN, Japan\\
  $^3$National Institute of Advanced Industrial Science and Technology (AIST), Japan\\
  $^4$T\'{e}l\'{e}com Paris, Institut Polytechnique de Paris, France
}
\email{
du@sap.ist.i.kyoto-u.ac.jp, 
\{adityaarie.nugraha, kouhei.sekiguchi\}@riken.jp, 
y.bando@aist.go.jp, mathieu.fontaine@telecom-paris.fr, yoshii@i.kyoto-u.ac.jp
}

\newcommand{\R}{\R}
\newcommand{\sscm}{\V}

\newcommand{\eg}{\textit{e.g.}}
\newcommand{\etal}{\textit{et~al.}}

\def\hr{\mathsf{H}}

\def\Tr{\mathrm{Tr}}

\def\R{\mathbf{R}}
\def\S{\mathbf{S}}

\def\V{\mathbf{V}}

\def\X{\mathbf{X}}

\def\Z{\mathbf{Z}}

\def\a{\mathbf{a}}

\def\e{\mathbf{e}}

\def\i{\mathbf{i}}

\def\s{\mathbf{s}}
\def\t{\mathbf{t}}
\def\u{\mathbf{u}}
\def\w{\mathbf{w}}
\def\x{\mathbf{x}}
\def\y{\mathbf{y}}

\begin{document}

\setlength{\intextsep}{5pt} %
\setlength{\textfloatsep}{5pt} %
\setlength{\abovecaptionskip}{5pt}
\setlength{\belowcaptionskip}{5pt}

\setlength{\abovedisplayskip}{3pt}
\setlength{\belowdisplayskip}{3pt}
\allowdisplaybreaks[4]

\maketitle
\begin{abstract}
\vspace{-1mm}
This paper describes noisy speech recognition for an augmented reality headset
 that helps verbal communication with in \textit{real} multiparty conversational environments.
A major approach that has actively been studied in \textit{simulated} environments
 is to sequentially perform speech enhancement and automatic speech recognition (ASR)
 based on deep neural networks (DNNs) trained in a supervised manner.
In our task, however,
 such a pretrained system fails to work
 due to the mismatch between the training and test conditions
 and the head movements of the user.
To enhance only the utterances of a target speaker,
 we use beamforming based on a DNN-based speech mask estimator
 that can adaptively extract the speech components 
 corresponding to a head-relative particular direction.
We propose a semi-supervised adaptation method
 that jointly updates the mask estimator and the ASR model at run-time
 using clean speech signals with ground-truth transcriptions
 and noisy speech signals with highly-confident estimated transcriptions.
Comparative experiments using the state-of-the-art distant speech recognition system
 show that the proposed method significantly 
 improves the ASR performance.
\end{abstract}
\noindent\textbf{Index Terms}: 
speech enhancement, speech recognition, human-computer interaction, 
run-time adaptation.

\section{Introduction} \label{sec-intro}		%

Smart glasses or augmented reality (AR) headsets 
 (\eg, Microsoft HoloLens\footnote{https://www.microsoft.com/en-us/hololens})
 equipped with multimodal IO devices
 including cameras, microphones, AR displays, loudspeakers, 
 and inertial measurement unit (IMUs)
 are expected to become more powerful and compact
 and totally change a way of daily communication.
One of the most important applications of speech processing technology
 is to help a (hearing-impaired) user make verbal communication
 by enhancing the speech of a target speaker
 while showing its transcriptions using an AR technique.
Such a system could be extended
 for helping a user talk with foreign people in a cocktail party with loud noise
 by translating the speech of a target speaker.
The key ingredient of these systems
 is distant noisy speech recognition capable of working robustly
 regardless of acoustic conditions and user movements.
The typical approach to noisy speech recognition 
 using a microphone array is to sequentially perform 
 multichannel speech enhancement and automatic speech recognition (ASR)~\cite{heymann_neural_2016,erdogan_improved_2016}.

Beamforming has recently been the main choice for speech enhancement
 and its performance depends on the accuracy
 of estimating the spatial covariance matrices (SCMs) 
 or steering vectors of speech and noise.
One may select the SCMs corresponding to the source and noise directions
 from those prepared in advance
 (\eg, computed from the array geometry
 or measured in an anechoic room).
This strategy often fails
 due to the mismatch between the prepared and real SCMs.
To estimate the speech and noise SCMs at run-time,
 one can train a deep neural network (DNN)
 that predicts speech masks in the short-time frequency transform (STFT) domain
 \cite{heymann_neural_2016,erdogan_improved_2016}.
To enhance only the utterances of a particular speaker 
 regardless of speech overlaps in multiparty conversation,
 the mask estimator can be informed 
 of the speaker direction and/or identity~\cite{8639593,8268910, Nakagome2020, Subramanian2020,Li2019,li_icassp2022,shao_icassp2022,subramanian_icassp2021}.
The performance of such supervised learning, however, is still severely limited
 due to the mismatch between the training and test conditions.
When an AR headset is used,
 the head movement of the user adversely affects the performance.
Existing methods of separating moving sources~\cite{1200008,5496044,8114273}
 are vulnerable to fast movements.

\begin{figure}[t]
  \centering
  \includegraphics[width=.8\columnwidth]{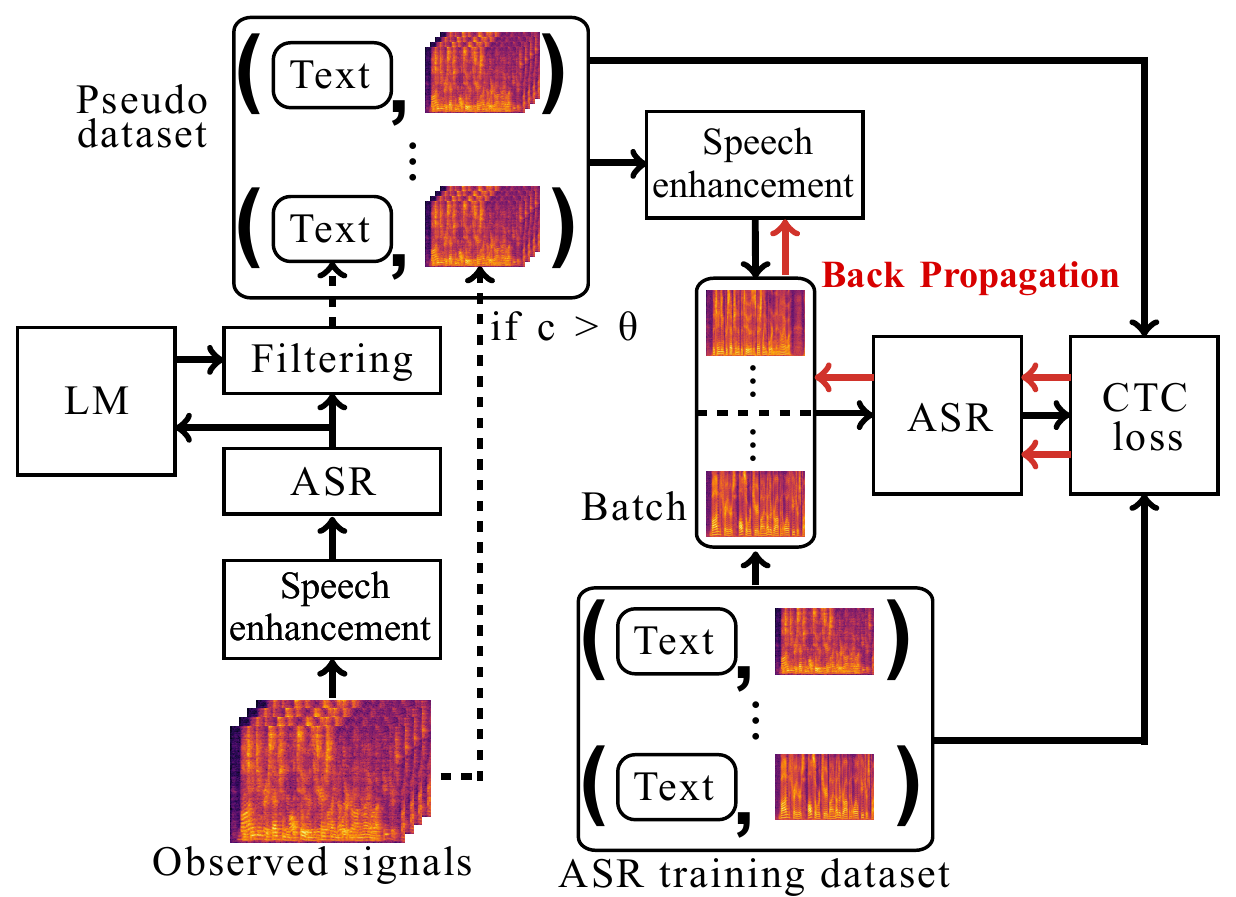}
  \vspace{-1mm}
  \caption{The proposed joint adaptation approach.}
  \vspace{-2.5mm}
  \label{fig:domain_adaptation}
\end{figure}

The condition mismatch
 has also been one of the central problems in 
 the state-of-the-art end-to-end approach to ASR 
 that directly maps speech signals to symbols (\eg, characters and words)
 with the attention mechanism 
 and/or the connectionist temporal classification (CTC) strategy.
The most basic way of improving the robustness against noise
 is multi-condition training of an ASR model~\cite{multiconditionASR2015, multiconditionASR2019}.
In multichannel speech recognition,
 some studies concatenate a speech mask estimator and an ASR model 
 in a differentiable manner
 such that both DNNs can be jointly optimized~\cite{Ochiai2017a,Subramanian2020}.
Most studies on these topics have been conducted under offline closed conditions
 where synthesized noisy speech signals with the ground-truth transcriptions
 are used for training and evaluation.
This prohibits the practical application of ASR to an AR headset in the real world.

Under these circumstances,
 we propose a speech recognition system for an AR headset
 that works robustly and adaptively in real conversational environments
 (Fig.~\ref{fig:domain_adaptation}).
As in the standard architecture,
 our system consists of speech enhancement and recognition modules,
 which are implemented 
 with minimum variance distortionless response (MVDR) beamforming 
 with DNN-based mask estimation
 and an audio-to-word CTC model, respectively.
These DNNs are separately trained in a \textit{supervised manner in advance}
 and jointly updated with backpropagation in a \textit{semi-supervised manner at run-time}
 using clean speech signals with ground-truth transcriptions taken from the training data
 and observed noisy speech signals with estimated transcriptions
 whose confidence scores are larger than a threshold.
To focus on a particular speaker in multiparty conversation,
 the mask estimator is given as directional clues 
 the premeasured steering vectors corresponding 
 to the head-relative speaker direction,
 which is assumed to be estimated by multimodal sensors.

The major contribution of this work
 is to enable both the speech enhancement and recognition modules
 to be jointly updated for run-time adaptation in real online environments
 unlike existing studies restricting the joint optimization 
 to the training phase in simulated offline environments~\cite{Ochiai2017a}.
We experimentally prove the importance of the run-time adaptation 
 on the recently-released Easy Communications (EasyCom) dataset~\cite{donley2021easycom}
 consisting of real multiparty conversation recordings.

\section{Related Work}
\label{sec:related_work}
This section briefly reviews
 how to reflect the characteristics of observed data
 for adapting beamforming-based speech enhancement
 and end-to-end speech recognition.

\subsection{Adaptive Speech Enhancement}
\label{sec:conventional_beamformer}

We review speech enhancement in the short-time Fourier transform (STFT) domain
 spanned by $F$ frequency bins and $T$ time frames.
Our goal is to estimate 
 the STFT coefficients of clean speech %
 $\S \!\triangleq\! \{s_{ft}\}_{f,t=1}^{F,T} \!\in\! \mathbb{C}^{FT}$
 from those of an observed multichannel noisy speech 
 $\X \!\triangleq\! \{\x_{ft}\}_{f,t=1}^{F,T} \!\in\! \mathbb{C}^{FTM}$ obtained from $M$ sensors.
In general, 
 $\x_{ft} \!\in\! \mathbb{C}^{M}$ 
 is assumed to be the output of a time-frequency linear system given by
\begin{align}
  \x_{ft} = \a_f s_{ft} + \e_{ft}, \label{eq:mixModel_timeInvA}
\end{align}
where
$\e_{ft} \in \mathbb{C}^{M}$ is additive noise and
$\a_f \in \mathbb{C}^M$
 is the steering vector of frequency $f$
 corresponding to the speaker direction.
The classic beamforming technique
 estimates the clean speech %
as
\begin{align}
  \hat{s}_{ft} 
  = \w_f^\hr \x_{ft} 
  = \w_f^\hr \a_f s_{ft} + \w_f^\hr \e_{ft},
\end{align}
where $\w_f \!\in\! \mathbb{C}^M$ is a demixing filter of frequency $f$
 and $^\hr$ is the conjugate transpose.
The
 minimum variance distortionless response (MVDR) beamforming \cite{souden_optimal_2010}
 computes $\w_f$ as
\begin{align}
  \w_f %
  = \R_f^{-1} \V_f \qty[\Tr \qty(\R_f^{-1} \V_f)]^{-1} \u,
  \label{eq:mvdr2}
\end{align}
where
 $\V_f \!=\! \mathbb{E}[\s_{ft}\s_{ft}^\hr]$ and 
 $\R_f \!=\! \mathbb{E}[\e_{ft}\e_{ft}^\hr]$ are the SCMs of the speech and noise, respectively,
 and $\u \in \{0, 1\}^M$ is a one-hot vector 
 indicating the reference microphone.

\subsubsection{Mask-Informed Beamforming}

Given soft masks %
 $\Z \!\triangleq\! \{z_{ft}\}_{f,t=1}^{F,T} \!\in\! [0, 1]^{FT}$,
 where $z_{ft}$ represents the ratio of the speech component
 at frequency $f$ and time $t$, 
 the speech and noise SCMs are typically computed as
$\V_f \!=\! \mathbb{E}\!\left[z_{ft} \x_{ft} \x_{ft}^\hr\right]$ and
$\R_f \!=\! \mathbb{E}\!\left[(1 - z_{ft}) \x_{ft} \x_{ft}^\hr\right]$, respectively.
In general, 
 the speech masks $\Z$ can be estimated 
 with a DNN that takes the noisy speech spectrogram $\X$ as input
 \cite{heymann_neural_2016,erdogan_improved_2016}.
Such a DNN is trained in advance such that the binary cross entropy 
 between the estimated masks $\Z$ and
 the ground-truth binary masks 
 is minimized.
This technique has been shown to achieve good performance
 in simulated environments
 because $\V_f$ and $\R_f$ can be \textit{adaptively} estimated from $\X$.
However,
 the DNN used for estimating $\Z$ is still \textit{fixed}
 and thus suffers from the condition mismatch problem
 in real environments.

\subsubsection{Direction-Aware Mask Estimation}

Several attempts have been made 
 for informing the mask estimator of directional information.
Li~\etal~\cite{Li2019} proposed an attention network 
 that estimate the speech spectra as the weighted sum of multiple fixed beamformers.
The enhanced spectra and an utterance of the target speaker
 are then fed to the SpeakerBeam network~\cite{8268910}
 that estimates the speech masks corresponding to the speaker direction and identity.
Chen~\etal~\cite{8639593} and Subramanian~\etal~\cite{subramanian_far-field_2020}
 computed directional features
 from the observed noisy speech spectra
 and the steering vectors of the target direction premeasured in an ideal environment.
These features are fed to a mask estimator
 together with the other common features 
 including inter-microphone phase differences and the output of a fixed beamformer.
The performances of these methods in real environments with unseen speakers
 would also be drastically degraded by the condition mismatch problem.

\subsection{Adaptive Speech Recognition}

The end-to-end approach to ASR
 has recently become popular
 due to the easy implementation, good performance, and fast inference.
Among various end-to-end models implemented
 with the attention mechanism~\cite{attention},
 the connectionist temporal classification (CTC) architecture~\cite{graves_connectionist_2006},
 the recurrent neural network (RNN) transducer~\cite{DBLP:journals/corr/abs-1211-3711},
 and the transformer~\cite{gulati20_interspeech},
 those based on the CTC and RNN transducer 
 have been experimentally shown to work relatively well
 in real noisy reverberant environments of the CHiME-6 Challenge \cite{watanabe20b_chime},
 but still have large room for improvement~\cite{andrusenko_towards_2020}.
In fact,
 even the state-of-the-art models 
 that achieve the best performance in standard ASR benchmarks fail to work
 for noisy distant ASR
 when the speaker characteristics or acoustic environments significantly differ
 between the training and test conditions.
Such overfitting to a particular dataset
 is one of the most serious problems in the ASR field.

The condition mismatch problem
 can be mitigated by semi-supervised adaptation methods
 that use annotated training data (clean speech) with ground-truth transcriptions
 and non-annotated test data (noisy speech) with the estimated transcriptions (pseudo-labels) 
 for further updating ASR models
 \cite{weninger20_interspeech,DBLP:journals/corr/abs-2010-15653,
 xu20b_interspeech,park20d_interspeech}.
To avoid using test data with erroneous transcriptions 
 for effective model adaptation,
 one may use a dropout-based uncertainty measure~\cite{DBLP:conf/icassp/KhuranaMHR21}.
In momentum pseudo labeling (MPL),
 the pseudo-labels are iteratively updated based on an ensemble of ASR models 
 at different time steps~\cite{higuchi21_interspeech}.

\section{Proposed Method}
\label{sec:proposed_method}
This section
describes the proposed method
integrating
a direction-aware mask estimator to solve the speech overlap problem,
a head-movement-aware speech enhancement to deal with the user's head movements,
and a semi-supervised environment adaptation algorithm
to adjust the enhancement and recognition model parameters to the test environment.

\subsection{Problem Specification}

We aim to estimate the clean speech $\S$
given the observed multichannel noisy speech $\X$.
Instead of using the time-invariant steering vector as in Eq.~\eqref{eq:mixModel_timeInvA},
we let the steering vector to be time-varying to deal with the user's head movements as follows:
\begin{align}
  \x_{ft} = \a_{ft} s_{ft} + \e_{ft}, \label{eq:mixModel_timeVarA}
\end{align}
where
$\a_{ft} \in \mathbb{C}^M$
is the time-varying steering vector corresponding to the speaker direction at time $t$.
A time-varying filter $\w_{ft} \in \mathbb{C}^M$ is thus required to perform speech enhancement.

\subsection{Direction-Aware DNN-Based Mask Estimator} \label{sec:mask_estimator}

Let $\mathcal{M} \!\triangleq\! \{1, \ldots, M\}$ 
 be the set of microphone indices,
 and $r$ be the index of a reference microphone.
The mask estimator takes as input
 spectral features, inter-channel phase difference (IPD) features, and directional features.
The spectral features are the logarithmic power spectra of $\X$ at the reference microphone,
 denoted by ${\{x_{ftr}\}_{f,t=1}^{F,T}}$.
The IPD features are the sines and cosines 
 of ${\{\{P_{ftm}\}_{f,t=1}^{F,T}\}_{m \in \mathcal{M} - \{r\}}}$,
 where $\smash{P_{ftm} \!\triangleq\! \angle \left(x_{ftm}/x_{ftr}\right)}$ 
 is the IPD between the $m$-th microphone and the reference microphone 
 at frequency $f$ and time $t$.
Similarly, the directional features are the sines and cosines 
 of $\smash{\{\{D_{ftm}\}_{f,t=1}^{F,T}\}_{m \in \mathcal{M} - \{r\}}}$,
 where $\smash{D_{ftm} \!\triangleq\! \angle \left(a_{ftm}/a_{ftr}\right)}$ is
 computed from the premeasured steering vectors of the target direction
determined by leveraging the multimodal sensors,
\eg, by applying a face detection algorithm on the camera images.
The mask estimator thus has $(F, T, 4 \!\times\! (M \!-\! 1) \!+\! 1)$-dimensional input
 and outputs time-frequency masks $\Z$ for the target speaker 
 for calculating the demixing filters.

\subsection{Head-Movement-Aware Speech Enhancement} \label{sec:head_movement}

To deal with the head movements of the user, %
we compute demixing filters for the time-varying target direction.
Let $D$ be the number of directions with 
$d \!\in\! \{1, \ldots, D\}$ is the index.
The MVDR demixing filter for direction $d$ is given by
\begin{align}
  \w_{fd} = \qty(\R_{fd})^{-1}\sscm_{fd} \qty[\Tr \qty(\qty(\R_{fd})^{-1}\sscm_{fd})]^{-1}\u ,
\end{align}
where
 the speech and noise SCMs of the direction $d$ are respectively given by
$\sscm_{fd} \!\triangleq\! 
    \sum_{t=1}^T \phi_{d,t} z_{ft} \x_{ft} \x_{ft}^\hr$ and
$\R_{fd} \!\triangleq\! 
    \sum_{t=1}^T \phi_{d,t} (1 - z_{ft}) \x_{ft} \x_{ft}^\hr$ with
$\phi_{d,t} = 1$ if the target speaker is in direction $d$ at time $t$
and $\phi_{d,t} = 0$ otherwise,
assumed to be known by utilizing multimodal sensors.
The separated signal $\hat{s}_{ft}$ is then obtained by
\begin{align}
  \hat{s}_{ft} = \sum_{d=1}^D 1_{d,t} \w_{fd}^\hr \x_{ft}.
\end{align}

\subsection{Adaptation of Speech Enhancement and Recognition}
\label{sec:domain_adaptation}

The performances of
the mask estimator trained on simulated noisy speech signals and
the ASR system trained on clean speech signals
are degraded in real noisy environments.
To address this environment mismatch issue,
we propose to update the mask estimator and parts of the ASR system
using transcription estimates that are deemed to be reliable
(Fig.~\ref{fig:domain_adaptation}).
The reliability is determined by a confidence score $c$ computed as 
\begin{align}
  c = \alpha \log p_\mathrm{ASR}(\y | \x) + \beta \log p_\mathrm{LM}(\y) + \gamma \qty|\x|,
  \label{eq:filtering}
\end{align}
where $p_\mathrm{ASR}(\y | \x)$ is the probability for the ASR system
to output sequence $\y$ given the enhanced signal $\x$,
$p_\mathrm{LM}(\y)$ is the the probability for the language model to output sequence $\y$,
$|\x|$ is the length of $\x$,
and $\alpha, \beta, \gamma \in \mathbb{R}$
are weight coefficients.
The pairs of observed signal and transcription estimate
whose confidence score exceeds a threshold $\theta$
are included in a pseudo dataset
that is used together with the original ASR training dataset to perform the adaptation.
It updates the parameters of the mask estimator and parts of the ASR system 
by minimizing a loss function:
\begin{align}
  \mathcal{L}^\text{AD} = \mathrm{CTCLoss}(\y, \t)
  + \lambda \qty||\boldsymbol{\omega} - \boldsymbol{\omega}^\mathrm{init}||^2,
  \label{eq:domain_adaptation_obj}
\end{align}
where $\y$ is the ASR output sequence and
$\t$ is
the ground-truth transcription for the batch sample from the original ASR training dataset or the transcription estimate for that from the pseudo dataset.
The second term of Eq.~\eqref{eq:domain_adaptation_obj}
regularizes the mask estimator updates,
where
$\boldsymbol{\omega}$ is the current parameters,
$\boldsymbol{\omega}^\mathrm{init}$ is the initial parameters before the adaptation,
and $\lambda \in \mathbb{R}$ is a weight.
Conversely, the regularization for the ASR system updates is achieved by the use of the original ASR training dataset.

\section{Evaluation}
\label{sec:evaluation}

This section reports our experiments
to validate the effectiveness of
the proposed joint adaptation of speech enhancement and recognition for an AR headset used in real environments.

\subsection{Experimental Settings} \label{sec:exp_settings}

We used the Easy Communications (EasyCom) dataset~\cite{donley2021easycom} for evaluation.
There are 12 sessions of conversational recordings (5 h 18 min)
 with the ground-truth transcriptions.
We used Session 2 to 12 for adaptation 
 and Session 1 for evaluation.
Although the dataset contains 6-channel audio data,
 we used only the first 4 microphones fixed on the AR headset.
All audio signals were resampled from 48 kHz to 16 kHz.
For dereverberation,
 we applied weighted linear prediction (WPE)~\cite{yoshioka_generalization_2012} 
 with a tap length of 16 and a delay of 2 %
 in the STFT domain with a window length of 512 pts and a shifting interval of 128 pts.
Using the provided impulse responses recorded in an anechoic chamber,
 we computed steering vectors for $D \!=\! 1020$ directions. 
At run-time, assuming the target direction to be accurately estimated from multimodal data,
 we exploited the user and target poses tracked with an OptiTrack system
 for giving the steering vectors of the target direction
 to the mask estimator.

Dividing the test data (Session 1) into non-overlapped and overlapped speech regions
 based on the voice activity annotation,
 we evaluated the impacts of the proposed head-movement-aware enhancement and environment adaptation
 in each region in terms of
 the word error rate (WER) and
 the source-to-distortion ratio (SDR) \cite{vincent_performance_2006},
 where the close microphone signal was regarded as the reference.
We obtained the baseline performance 
 by feeding the observed signals of the first microphone to the ASR system without enhancement.
Similarly, we also obtained the oracle performance 
 by feeding the signals of the close microphone.
The rest of this subsection %
 details the settings for the mask estimator, 
 the ASR system, including the language model, and the environment adaptation.

\subsubsection{Mask Estimator Settings} \label{sec:mask_system}

The mask estimator was trained on simulated data generated using
the room simulation in Pyroomacoustics~\cite{scheibler_pyroomacoustics_2018}.
The simulated data mimics the scenario in the EasyCom dataset,
where several people make conversation while sitting around a table.
Nonetheless, we varied the rooms by randomly sampling
the width from $[5.0\,\text{m}, 7.0\,\text{m}]$, 
the depth from $[6.0\,\text{m}, 8.0\,\text{m}]$, 
and the height from $[2.5\,\text{m}, 3.5\,\text{m}]$
with a reverberation time sampled from $[150\,\text{ms}, 300\,\text{ms}]$.
The device user positions w.r.t. the room dimensions were sampled from
$[0.4 \!\times\! \text{width}, 0.6 \!\times\! \text{width}]$,
$[0.15 \!\times\! \text{depth}, 0.35 \!\times\! \text{depth}]$, and 
$[1.0\,\text{m}, 1.5\,\text{m}]$, respectively.
We assumed the device had a 4-microphone array
 whose geometry followed the one in the EasyCom dataset.
Let the azimuth and elevation of the positive depth direction be $0.0^\circ$.
For each utterance, the azimuth and elevation of the device were initially sampled from 
$[-72^\circ, 72^\circ]$ and $[-45^\circ, 45^\circ]$, respectively,
and resampled once more from the same ranges to simulate a head movement.
The target speaker and interferer positions w.r.t. the room dimensions were sampled from
$[0.1 \!\times\! \text{width}, 0.9 \!\times\! \text{width}]$,
$[0.4 \!\times\! \text{depth}, 0.85 \!\times\! \text{depth}]$, and 
$[1.0\,\text{m}, 1.5\,\text{m}]$, respectively.

The utterances of the target speaker and interferer were taken 
 from the train-clean-100 subset of the LibriSpeech corpus \cite{panayotov_librispeech_2015}.
For each target speaker's utterance, the interferer was randomly active with a probability of 50\%.
A background noise taken from the CHiME-3 dataset \cite{barker_third_2015} was then added to the simulated multichannel signals such that the signal-to-noise ratio (SNR) was in $[-2\,\text{dB}, 8\,\text{dB}]$.
In total, the training dataset contained 28,540 pairs of
multichannel mixture signal and its corresponding ground-truth target speech signal. 

The mask estimator consisted of three bi-directional LSTM layers with 256 hidden units,
 a dropout layer with a dropout rate of 0.2, 
 and a fully connected layer with 201 units.
It was trained by minimizing
the phase-sensitive magnitude spectral approximation loss \cite{erdogan_phase-sensitive_2015} given by
\begin{align}
  \mathcal{L}^\text{MA} \!=\!
  \sum_{f=1}^F \sum_{t=1}^T
  \norm{z_{ft}|x_{ft}| - |s_{ft}|\max\!\qty(0, \cos \Delta_{ft})}^2,
\end{align}
where
$|\!\cdot\!|$ computes the magnitude and
$\Delta_{ft} \!\triangleq\! \angle x_{ft} \!-\! \angle s_{ft}$
with
$\angle$ computes the phase. 
It was trained for 200 epochs
using the AdamW optimizer \cite{loshchilov_decoupled_2019} with
$\beta_1 \!=\! 0.9$, $\beta_2 \!=\! 0.999$, and a weight decay of $0.01$.
The learning rate was initially set to $0.001$ and then decayed for $0.96$ every epoch.

\subsubsection{ASR System Settings} \label{sec:asr_system}

We used a CTC-based \cite{graves_connectionist_2006} end-to-end ASR model
consisting of
2 VGG-like convolutional layers,
6 bi-directional LSTM layers with 512 hidden units,
and 1 fully connected layer.
The input features were 80-dimensional log-mel spectrograms
computed with a window length of 400 samples (25 ms)
and a shift interval of 160 samples (10 ms).
The outputs were 8191 subwords tokenized by byte pair encoding (BPE)~\cite{sennrich-etal-2016-neural}.

The ASR model was pretrained on the LibriSpeech corpus \cite{panayotov_librispeech_2015}.
It was then trained on 
the CHiME-6 dataset \cite{watanabe20b_chime},
the CommonVoice corpus \cite{commonvoice:2020},
and a noisy reverberant dataset we derived from the CommonVoice corpus
by convolving the clean utterances with simulated room impulse responses
and adding background noise taken from the CHiME-3 dataset \cite{barker_third_2015}
such that the SNR is in $[5\,\text{dB}, 10\,\text{dB}]$.
The pretraining and training were performed for 100 epochs
using the AdamW optimizer~\cite{loshchilov_decoupled_2019}
as for the mask estimator.
The learning rate was
first linearly increased from $0$ to $1.5 \times 10^{-4}$ over the minibatches in the first epoch,
and then decayed for $0.97$ every epoch.
The minibatch size was $32$ and the dropout rate was $0.2$.

The language model used for
computing $\log p_\mathrm{LM}(\y)$
was a Transformer-based model~\cite{NIPS2017_3f5ee243}
composed of 
4 encoder layers with 8-head attention and 1 fully connected sublayers,
whose output dimensions are 512.
It was trained on the CommonVoice and the CHiME-6 datasets
for 10 epochs
using the Adam optimizer
with varying learning rate as in \cite[Sec. 5.3]{NIPS2017_3f5ee243}.

\subsubsection{Environment Adaptation Settings}

Based on preliminary experiments,
the coefficients in Eqs. \eqref{eq:filtering} and \eqref{eq:domain_adaptation_obj} are set to
$\alpha \!=\! 1$, $\beta \!=\! 50$, $\gamma \!=\! 1000$, and
$\lambda \!=\! 5 \!\times\! 10^{-4}$.
The whole mask estimator
and the convolutional layers of the ASR system
were updated for 20 epochs
using the Adam optimizer \cite{DBLP:journals/corr/KingmaB14} with $\beta_1 \!=\! 0.9$, $\beta_2 \!=\! 0.999$, and
a learning rate of $5 \!\times\! 10^{-4}$.
One minibatch was composed of
32 samples from the original ASR training dataset
and 32 random samples from the pseudo dataset.
To investigate the impacts of adaptation data amount,
we varied the threshold $\theta$ such that
the top 250, 500, 1000, 1500, or 2363 (of 2363) utterances with the highest confidence scores are used as the pseudo dataset.
The pseudo dataset was reconstructed every 5 epochs.
For comparison, we also investigate an oracle setting,
where the ground-truth transcriptions are used for the adaptation.

\subsection{Experimental Results and Discussion}

\begin{table}[t]
  \captionsetup[subfloat]{captionskip=0mm}
  \caption{
    The average performances obtained
    with/without the proposed head-movement awareness (denoted by `HMA') 
    and environment adaptation using different numbers of transcription estimates.
    * indicates the use of ground-truth transcriptions.
  }
  \vspace{-.2\baselineskip}
  \centering
  \subfloat[Non-overlapped Speech]{
    \setlength\tabcolsep{6pt}
    \setlength\aboverulesep{1pt}
    \setlength\belowrulesep{1pt}
    \begin{tabular}{ccccc}
      \toprule
      MVDR & HMA & Adaptation & SDR [dB] & WER [\%] \\ 
      \midrule
      - & - & - & -1.28 & 55.53 \\
      \checkmark & - & - & 0.64 & 47.45 \\
      \checkmark & \checkmark & -  & 1.99 & 41.99 \\
      \checkmark & \checkmark & 250 & 2.12 & 39.25 \\
      \checkmark & \checkmark & 500 & 2.10 & 38.39 \\
      \checkmark & \checkmark & 1000 & 2.18 & 38.38 \\
      \checkmark & \checkmark & 1500 & 2.10 & 39.38 \\
      \checkmark & \checkmark & 2363 & 2.13 & 44.10 \\
      \checkmark & \checkmark & 
      2363*
      & 2.11 & 38.39 \\
      \multicolumn{3}{c}{Oracle (close mic.)} & $\infty$ & 24.72 \\
      \bottomrule
    \end{tabular}}\\[-3mm]
  \subfloat[Overlapped Speech]{
    \setlength\tabcolsep{6pt}
    \setlength\aboverulesep{1pt}
    \setlength\belowrulesep{1pt}
    \begin{tabular}{ccccc}
      \toprule
      MVDR & HMA & Adaptation & SDR [dB] & WER [\%] \\ 
      \midrule
        - & - & -  & -6.30 & 81.70 \\
        \checkmark & - & - & -9.46 & 78.06 \\
        \checkmark & \checkmark & -  & -5.30 & 71.29 \\
        \checkmark & \checkmark & 250 & -4.97 & 64.87 \\
        \checkmark & \checkmark & 500 & -4.93 & 63.23 \\
        \checkmark & \checkmark & 1000 & -4.76 & 62.36 \\
        \checkmark & \checkmark & 1500 & -5.44 & 63.75 \\
        \checkmark & \checkmark & 2363 & -5.24 & 74.85 \\
        \checkmark & \checkmark & 2363* & -3.14 & 57.94 \\ 
        \multicolumn{3}{c}{Oracle (close mic.)} & $\infty$ & 33.39 \\
      \bottomrule
    \end{tabular}}
  \label{table:domain_adaptation_result}
  \vspace{-1mm}
\end{table}

Table~\ref{table:domain_adaptation_result} shows
the effectiveness of the proposed environment adaptation
for both non-overlapped speech and overlapped speech.
It indicates that the proposed joint adaptation of neural speech enhancement and recognition 
alleviates the mismatch between the acoustic properties of real environments
and those of simulated environments.
However, the threshold $\theta$ for the pseudo dataset construction
must be adjusted appropriately
to avoid a detrimental effect on the performance.
In our experiments, we found that using the top 1000 (of 2363) utterances with the highest confidence scores was optimal.

There was still a huge gap between the performance of our proposed system and the performance of the oracle setting.
For the overlapped speech case,
the WERs of the proposed system were 71.29\% before adaptation
and 62.36\% after adaptation, while
the WER of the oracle setting using the close microphone was 33.39\%.
We observed that the device user's speech was not suppressed well due to inaccurate mask estimates.
It can be attributed to the fact
that the user's speech was captured much louder than the target's %
because the user's mouth was closer to the microphones and
that the spatial characteristics of the user's speech were significantly different from those of training data.

\section{Conclusion}
\label{sec:conclusion}

This paper presented a joint adaptation method
 for a neural speech enhancement and recognition system
 to handle the data mismatch problem related to unseen environments or speakers.
The method exploits a number of transcription estimates 
 with higher confidence scores for run-time model updates,
 in which the original training data are also used for regularization.
For the evaluation, we considered its application
 to the ASR system for an AR headset 
 operated in real conversational environments.
Our experiments showed that
the proposed joint adaptation improved
the WERs by 3.61 points for the non-overlapped speech case
and by 8.93 points for the overlapped speech case.
Future work includes better treatment of the user's own speech.

\section{Acknowledgments}

This work was supported in part 
by JSPS KAKENHI Nos.~19H04137, 20K19833, and 20K21813.

\clearpage
\bibliographystyle{./IEEEtran}
\bibliography{./IEEEabrv,./MYabrv,./references}

\end{document}